# THE SURVEY OF SENTIMENT AND OPINION MINING FOR BEHAVIOR ANALYSIS OF SOCIAL MEDIA

Saqib Iqbal[1] and Ali Zulqurnain[1], Yaqoob Wani[1] and Khalid Hussain[2]

[1]School of Information Technology, University of Lahore, Pakistan
[2]Faculty of Computing, Universiti Teknologi Malaysia, Malaysia

## ABSTRACT

*Nowadays, internet has changed the world into a global village. Social Media has reduced the gaps among the individuals. Previously communication was a time consuming and expensive task between the people. Social Media has earned fame because it is a cheaper and faster communication provider. Besides, social media has allowed us to reduce the gaps of physical distance, it also generates and preserves huge amount of data. The data are very valuable and it presents association degree between people and their opinions. The comprehensive analysis of the methods which are used on user behavior prediction is presented in this paper. This comparison will provide a detailed information, pros and cons in the domain of sentiment and opinion mining.*

## 1. INTRODUCTION

Social media has become an essential part of life, it is a comparatively cheap and widely accessible medium that enables anyone to broadcast and access information / news / knowledge, and to build new relations. It is a tool which is used to share different opinions that can belong to different subjects; such as humanitarian causes, environmental irregularities, economic issues, or political disputes. Honeycomb Framework describes social media by using seven functional building blocks: identity, conversations, sharing, presence, relationships, reputation, and groups[1].

Threaded deliberations, individuals not only express their views on an issue but also interact with one another. However, the discussions could become extremely emotional and intense with many expressive sentiments among individuals.

Microblogging (Twitter or Facebook) is the most popular communication channel among people in this era. Many researchers are working on sentiment and opinion mining on textual information which is gathered from the microblogging platforms. These sentiments and opinions are being used to improve business's marketing and productivity. Researchers employ the usage of several mining techniques in order to extract the sentiments and to improve the sentiments extractions results from textual, structured and un-structured formats.

Twitter (microblogging) revolution which occurred during Arab springs has raised two issues which are now being discussed frequently i.e. the uses of Twitter (microblogging) for publicizing local causes to distant audiences, and its importance in logistical communication among protesters on the ground[2].

Although sentiment and opinion mining is very helpful in order to draw the behavior of users against products, movies or events. More work is needed in order to grow the social media such that it can allow individuals to monitor and predict any uncertainty, and how to address and cope with such situations. Sentiment and opinion mining can become a more powerful tool by improving and addressing how many users support a sentiment whether it is in negation or in favor about a topic, and to identify the like-minded cluster of people and their relationship through sentiment mining.

## 2. RELATED WORK

Nidhi Mishra *et al*., 2012has described the importance of opinion and sentiments mining. Opinion mining is distributed into four sub-categories: sentence level, document level, feature level and compound level respectively, and has discussed the various tools and mining methods like precision, recall and F-measure on movie reviews or product reviews. Aspect level classification can provide greater detail for analysis then document and sentence level. The distribution categories miss classification subjective expressions and objective expressions for better understanding of sentiments [3,4,5].

Meeyoung Cha *et al*., 2010investigated the dynamics of public influence across different issues / topics to determine the influence author used in degree, retweets and mentions approaches on a large dataset of twitter. Spearman's rank correlation coefficient has been used for evaluating the influence of people against three different topics. Author concludes the result in terms of influence against each topic but he didn't target the fake users which can actively participate in discussions. Fake user can play the major role to divert the influence of public from one issue to another. [6,7].

Guendalina*et al*., 2015presents a new approach known as the integrated mixed methods approach (IMiME) for monitoring the content of social media by demonstrating a case study 'hypoactive sexual desire disorder (HSDD). Integrated Mixed methods approach (IMiME) is the combination of qualitative and quantitative approaches. Although author has presented the result of IMiME approach as balanced between ecological and situated nature of qualitative research strategies with systematization of quantitative measurement. This approach have to be tested with other case studies for concrete satisfaction [8, 9, 10].

Rui Gaspar *et al*., 2014has assessed the ways of coping, expressed by public in social media (Twitter) during food crises for its management and prevention. Methodological, conceptual limitation and challenges are discussed for coping analysis and how to apply psychosocial process to cope the content of single tweet. Qualitative approach is used for cope analysis. Qualitative approach is much hard and time consuming to analyze large dataset. Author selected limited data after cleansing for better results. During selection of data, author didn't consider the spam opinions. Spam opinions can be more harmful for decision making. Qualitative approach is used, if the Quantitative approach may also be added than it will be helpful for coping analysis and supportive for decision making [11, 12, 13, 14].

JianshuWeng*et al*., 2010 explored the most influenced users of Twitter (micro-blogging site) by homophily. In homophily, a single user choose and follow a friend on twitter because he is interested in the topics which were published in tweets by the friend and friend follows back because he observes that they both have similar interest topics.  Author endorse the homophily is not being used in case of twitter by using TwitterRank Approach. TwitterRank approach is the combination of Latent Dirichlet Allocation (LDA) and TwitterRank algorithms. Topic-sensitive PageRank has also been discussed, and its comparison with TwitterRank has been put forth. A group of spam/fake users can add someone to become more influenced user like celebrities, politicians etc. The author didn't detect the group of fake users as optimistic approach [15, 16].

Federico Neri *et al*., 2012has applied the sentiment mining analysis on Facebook textual dataset for news channel by using Recall and Precision approach. In this approach, it is very difficult to identify what is relevant to one person might not be relevant to another person and vice versa. Naïve Bayes can be used instead of Precision and Recall for better results [17, 18].

Jayasanka *et al*., 2014has assessed the challenges which were observed during extraction of expressions of optimistic or pessimistic attitudes through social media and has also discussed the different available tools and their limitations for sentiment analysis. Naïve Bayes classification method which is used in order to produce better results of sentiment mining instead of Natural language Processing and SentiWordNet. Scores are used in order to project the results of negative and positive sentiments. Expectation-Maximization (EM) algorithm can also be used in order to optimize the accuracy as the SentiWordNet does not make much difference among short sentences like 'not good' and 'not bad' [19, 20].

Ion SMEUREANU *et al*., 2012has presented a work for exploring the positive and negative opinions on pre-classified movies' reviews. Naïve Bayes Algorithm was applied on collection of comments. Precision and recall methods are used for accuracy check of Naïve Bayes algorithm and its execution time [21, 22].

Tkalcic *et al*., 2014has presented the scenario and design of personalized intervention system (PIS) and has chosen theory of planned behavior (TPB) method to predict whether user will attend the classical musical concert or not through social media mining. As theory of planned behavior (TPB) does not consider the emotional elements like feelings, mode; emotion element which has direct effect on personalized intervention system (PIS) predictions is not considered[23].

Sergio Oramas., 2014has used the social media (micro-blogging sites) and web (Musical content sites) for Musical Information Retrieval (MIR). Social Media Mining, Knowledge Extraction and Natural Language Processing techniques are combined in Musical Information Retrieval (MIR). Musical Information Retrieval (MIR) will work on unstructured and structured data. Structured and unstructured data will get from web and micro-blogging sites respectively[24].

Ahmed Hassan *et al.,* 2010has presented a method in order to detect an individual's attitude in a sentence towards others in threaded base communication. This approach is a combination of supervised Markov model of text, part-of-speech and dependency patterns. Support Vector Machines (SVM) is used for experiment results and to compare with other baselines: which are accuracy, recall and precision and F1[25].

Robert McColl *et al*., 2014has performed a comparison of execution and performance with twelve(12) open source graph databases. The Author has chosen same set of hardware, nodes and edges for execution and performance. Each open source graph database is evaluated with each algorithm which are Single Source Shortest Path (SSSP), Connected Components, PageRank and Update Rate[26].

Pravesh *et al*., 2014has assessed and evaluated the numerous methods which are used for opinion and sentiment. Naïve Bays Classifier, Support Vector Machine (SVM), Multilayer Perceptron, Clustering techniques has been used in order to analyze and compare the results of each technique. Every technique offer various benefits and boundaries, each method can be utilized according to the situation for features and text extractions. There are many other methods which can be utilized for extraction of sentiments from micro-blogging sites[27].

International Journal of Computer Science & Engineering Survey (IJCSES) Vol.6, No.5, October 2015Anindya Ghose *et al*.,2007 has proposed two ranking mechanisms: consumer-oriented ranking, manufacturer-oriented mechanism. These mechanisms are consisted on econometric and subjectively analysis and text mining. Consumer-oriented mechanism is used in order to rank the reviews according to their expected helpfulness while manufacturer-oriented mechanism is used in order to rank the reviews according to their expected effect on sales. Reviews can be spam or genuine but in these mechanism no spam detection method has been used. If spam reviews are also included in the ranking system, it will spoil the results and detract the community [28].

Table-1: Comprehensive research Analysis of Sentiment and Opinion Mining

| Paper Ref. No | Techniques/Methods/Protocols | Advantages | Disadvantages | Remarks |
|---|---|---|---|---|
| [3] | 1. Document Level<br>2. Sentence level<br>3. Task of Opinion mining at Feature level<br>4. Opinion Mining in Compound sentence | Author classify the opinion into different levels for understanding and mining. | Document and sentence level mining does not provide the necessary detail needed for many other applications. | Aspect level Opinion mining should also include in classification for better understand |
| [6] | Spearman's rank coefficient | Author use quantitative approach and project the users influence against different subjects/objects | Didn't detect the Fake users and follows | Fake users can change the results |
| [8] | Integrated mixed Methods(IMiMe) approach | Mixed-methods approach is combination of different qualitative and quantitative analytical strategies to overcome the drawbacks of available methods/tools. | IMiMe approach is case specific and needed to be checked and verified with other cases | Understanding and prevention User behavior |
| [11] | Coping Analysis/ psychosocial processes | Determine Ways of coping from social media and discuss the methodological, conceptual limitations and challenges | Spam opinion detection is missing, which can become the cause of worse decision | Opinion Spam Detection still a challenging problem which need to be sort out |
| [15] | Latent Dirichlet Allocation (LDA) /TwitterRank | Author used Unsupervised learning and quantitative approach for track the most influencer users of Twitters | Group spammers can play a the major role to make someone the most influencer user, The detection of group spammer is missing | Group spammers can mislead the community and change the thought. |
| [17] | Recall and Precision | Author used Social media specifically Facebook for sentiment mining for marketing | Precision and Recall, difficult to identify what is relevant or irrelevant to one person may not be relevant/irrelevant to another | Naïve Bayes can be used instead Precision and Recall for better results. |

24

| | | | | |
|---|---|---|---|---|
| [18] | Naïve Bayes Classification ,Maximum Entropy, Natural Language Tool Kit (NLTK), SentiWordNet | Author discuss sentiment extraction challenges and apply different techniques on data. | SentiWordNet cannot analyze short terms words like 'not bad' or 'not good' which reduce the accuracy | Expectation-Maximization (EM) algorithm can use to optimize the accuracy |
| [21] | Naive Bayesian classifier | Author verify the efficiency of Algorithm on positive and negative on user reviews. | Naïve Bayes can't learn interactions between features | Accuracy of SVM is better than Naïve Bayes, must try SVM instead Naïve Bayes |
| [23] | Theory of Planned Behavior Model (TPB) | TPB model is used for predicting the behavior of users | theory of planned behavior overlooks emotional variables such as threat, fear, mood and negative or positive feeling | Theory of planned behavior is not useful to prediction because it doesn't consider emotions |
| [24] | Music Information Retrieval (MIR). | a methodology that combines Social Media Mining, Knowledge Extraction and Natural Language Processing techniques, to extract meaningful context information for music from social data. | Application will not distinguish between comparative sentiments and regular sentiments. | should also include comparative and regular sentences mining for accuracy in mining the textual |
| [27] | 1. Single Source Shortest Path (SSSP) 2. Connected Components 3. PageRank 4. Update Rate | Discuss the execution power and comparison between open Source Graph databases with same hardware and collection of edges. | Support vector machine can also be utilized for comparison between open source graph databases | Open Source Graph Database can help to analyses the like-minded group of people. |
| [25] | Markov model of text, part-of-speech, dependency patterns | it predicts whether a sentence contains an attitude toward a text recipient or not | Not all subjective sentences have opinions | Can implement on microblogging sites for analysis |
| [26] | 1. Naïve Bays Classifier 2. Support Vector Machine (SVM) 3. Multilayer Perceptron 4. Clustering | Author evaluate different algorithms of Opinion and sentiment mining and compare them | Precision and Recall technique can also compare for sentiment and opinion mining | User's comments can be used to draw like-minded people and their interests. |
| [28] | consumer-oriented ranking, manufacturer-oriented ranking | ranking mechanism is combination of econometric analysis, text mining and subjectivity analysis | Reviews can be Spam or not spam, it really effect on market for selling product | Spam reviews should not be included during ranking of any product |

## 3. DISCUSSION

Sentiment and opinion mining is a very crucial / ambiguous process for extraction and evaluation of the behavior of public sentiment from social media. In this paper we discussed different research papers regarding sentiment analysis and opinion mining presented in Table-1. Researchers utilize and evaluate different algorithms and approaches forsentiment mining on International Journal of Computer Science & Engineering Survey (IJCSES) Vol.6, No.5, October 2015different subjects, different areas of sentiment and opinion mining are targeted to improve mining process. This research-works is very beneficial for improving the analysis and behavior prediction.

Prime focus of this study is classification of sentiment, evaluate the efficiency and accuracy of different sentiment and opinion mining algorithms like SVM (Support Vector Machine), Naïve Bayes Classification, Latent Dirichlet Allocation (LDA) , PageRank, Markov model of text, Multilayer Perceptron and Clustering.

Social media is a massive source of communication, where every individual post their sentiments. Detection of bogus or fraudulent sentiments or reviews become very hard from these sentiments and reviews which can post through fake users and spammers. These bogus or fraudulent sentiment can become a powerful approach and helpful to divert the mindset of individuals to boost any product demand and market capturing. Social media support many native/national languages of different countries which make hard to analyze sentiment of different countries on same topic or event. Use of informal or casual language reviews or opinions is another major issue during sentiment analysis because understanding of causal or informal language is much difficult then the formal language.

## 4. CONCLUSION

In Social media, people usually follow various events such as natural disasters, political issues, sports events and posts there comments / sentiments about the particular event. By applying sentiment and opinion mining techniques, it can be deduced that whether the sentiment is in denial or in favor of the event. But the most important point is how many people in total support and broadcast this particular sentiment, and how to identify the relationship amongst the supporters of a specific sentiment. This approach will be able to predict any uncertainty which may be either harmful or beneficial for the institutions.

A detailed assessment of different papers / articles is presented in this paper. There are several work which has been done on sentiment and opinion mining of social media but this mining technique is currently unformed. There are various research gaps in sentiment and opinion mining which are explained in the work done by the researchers. Many researchers are still researching on sentiment and opinion mining and how to improve this particular mining techniques.This paper is based on a comprehensive study on sentiment and opinion mining, and discuss the limitations and address the new areas in the domain of sentiment and opinion mining.